\RequirePackage{fix-cm}

\documentclass[smallextended 
]{svjour3}
\smartqed
\usepackage{amsmath,amssymb,graphicx,subfigure}

\begin{document}

\title {Scalar field potentials for closed and open cosmological models}
\author {Alexander Y. Kamenshchik \and Serena Manti}
\institute {Alexander Y. Kamenshchik \at Dipartimento di Fisica and INFN, Via Irnerio 46,40126 Bologna,
Italy\\
L.D. Landau Institute for Theoretical Physics of the Russian
Academy of Sciences, Kosygin str. 2, 119334 Moscow, Russia\\
\email{Alexander.Kamenshchik@bo.infn.it} 
\and Serena Manti \at Dipartimento di Fisica,Via Irnerio 46, 40126 Bologna,
Italy\\
Scuola Normale Superiore, Piazza dei Cavalieri 7, 56126 Pisa, Italy\\
\email{serena.manti@sns.it}}

\maketitle
\begin{abstract}
We develop a technique for the reconstruction of the potential for a scalar field or a tachyon field, reproducing a given 
cosmological evolution in a closed and open isotropic cosmological models. Such potentials are explicitly written down for 
the cases of the evolutions driven  by a generic barotropic fluid and by radiation plus a cosmological constant, for the case of a scalar field. For tachyon and pseudo-tachyon fields the potentials are reconstructed for  some special cases, corresponding to particular values of the barotropic index. 
\end{abstract}

\section{Introduction}
\label{intro}
The technique of the reconstruction of the potentials for scalar fields reproducing a given cosmological evolution has attracted the attention of researchers for a long time \cite{Starobinsky:1998fr}--\cite{Guo:2005ata}. Indeed, in modern cosmology the study of scalar fields has become very important, since the latter are considered as possible candidates for the role of the inflaton field, responsible for the inflationary phenomenon in the early universe \cite{inflation}, and of the dark energy substance \cite{darkenergy}, which causes the recent acceleration in cosmic expansion \cite{acceleration}. The main part of the corresponding research was devoted to a 
standard scalar field minimally coupled to gravity.  The case of the scalar fields non-minimally coupled to gravity was considered in 
\cite{Boisseau:2000pr}. The case of two scalar fields was studied in \cite{Andrianov:2007ua}. The technique of reconstruction of potentials 
for the cosmological models based on the tachyon fields \cite{Sen} was presented in \cite{Padmanabhan:2002cp,Feinstein:2002aj,Gorini:2003wa}.
The case of induced gravity was studied in detail in \cite{Kam-Tron-Ven}. All the papers mentioned above were devoted to the study of the flat Friedmann universe. Strangely, at least up to our knowledge, the case of the closed and open Friedmann universe was not studied. The only exception is the paper \cite{Ellis}, where the possibility of the treatment of the closed and open cosmological models was mentioned and some simple examples were presented. 
Here we would like to fill this gap. 
In this paper we reconstruct the potentials for a scalar field or a tachyon field reproducing a given 
cosmological evolution in closed and open isotropic cosmological models. In section 2 we describe  the technique of the reconstruction of the potential both for a scalar field and a tachyon field. In section 3 we explicitly find the solutions for the scalar field potential which reproduces the cosmological evolutions driven  by a generic barotropic fluid and by radiation plus a cosmological constant. In section 4 we reconstruct  tachyon and pseudo-tachyon \cite{Gorini:2003wa} potentials for the cases of the evolutions driven by the Chaplygin gas \cite{we-Chap,Fabris}, by the  anti-Chaplygin gas \cite{Carter,Gorini:2003wa} and for other special cases, corresponding to particular values of the barotropic index ($w=-\frac{2}{3},-\frac{1}{3},\frac{1}{3}$).  Section 5  contains concluding remarks.

\section{Technique of potential's reconstruction}
\subsection{Scalar field}
To begin with let us generalize  the technique of the reconstruction of potentials for the Friedmann-Lema\^itre cosmological model with the metric 
\begin{equation}
ds^2 = dt^2-a^2(t)\left(\frac{dr^2}{1-kr^2}+r^2(d\theta^2 +\sin^2\theta d\phi^2)\right)
\label{Fr}
\end{equation}
filled with a  spatially homogeneous scalar field $\varphi (t)$ with potential $V(\varphi )$. 
Here, as usual, $k=-1,0,1$ correspond to  open, flat and closed cosmologicla models respectively.
The energy density and the pressure are  
\begin{equation}
\rho =\frac{\dot \varphi ^2}{2}+V(\varphi )
\label{energy}
\end{equation}
and
\begin{equation}
p=\frac{\dot \varphi ^2}{2}-V(\varphi ) \hspace{0.1cm},
\label{pressure}
\end{equation}
where the dot indicates the derivative with respect to the cosmic time $t$. 
The energy conservation condition is 
\begin{equation}
\dot{\rho }=-3H(\rho +p) \hspace{0.1cm},
\label{conserv}
\end{equation}
where the Hubble parameter is defined as 
\begin{equation}
H \equiv \frac{\dot{a}}{a}.
\end{equation}

To obtain the expression for the scalar field potential, we first find the solution for $\varphi $ in terms of the scale factor $a$.\\
Summing (\ref{energy}) and (\ref{pressure}) we obtain
\begin{equation}
\dot{\varphi }^2=\rho +p
\label{varphi-dot}
\end{equation}
and, using $\dot{\varphi }=\dot a\frac{d\varphi }{da}$ and the first Friedmann equation with $8\pi G/3 = 1$, which is  
\begin{equation}
H^2+\frac{k}{a^2}=\rho , 
\label{Fried}
\end{equation}
we can write
\begin{equation}
\frac{d\varphi}{da}=\pm \frac{1}{a}\sqrt{\frac{\rho +p}{\rho -\frac{k}{a^2}}}
\label{phi-int}
\end{equation}
where $\rho $ and $p$ are  functions of $a$; integrating (\ref{phi-int}) we  arrive at the following  solution for the scalar field:
\begin{equation}
\varphi _\pm (a)=\pm \int \frac{da}{a}\sqrt{\frac{\rho (a)+p(a)}{\rho (a) -\frac{k}{a^2}}} 
\label{phi-int1}
\end{equation}
Generally, to different signs in the right-hand side of Eq. (\ref{phi-int1}) correspond different potentials, however, sometimes these potentials coincide (see e.g. \cite{Ellis,Gorini:2003wa}).   
Now, using the relation
\begin{equation}
V=\frac{1}{2}(\rho -p)
\label{poten}
\end{equation}
for the scalar field potential, obtained by subtracting (\ref{pressure}) from (\ref{energy}), and substituting the expressions for $\rho $ and $p$ in terms of the scale factor, we obtain  an expression for $V$ in terms of $a$. Supposing that (\ref{phi-int1}) is invertible,   we obtain 
\begin{equation}
V=V(a)=V[a(\varphi )] \hspace{0.1cm}.
\end{equation}

\subsection{Tachyon and pseudotachyon}
The tachyon field cosmological models inspired by string theory were first suggested in \cite{Sen}. 
The tachyon models represent a subclass of the models with non-standard kinetic terms \cite{k-ess}, which descend from the Born-Infeld model, invented already in the thirties \cite{Born}.
The relation between the models with non-standard kinetic terms and the relativistic hydrodynamics was studied in detail in 
paper \cite{hydroK}.
Being in some sense the simplest subclass of the general class of models with non-standard kinetic terms, the tachyon models 
have rather non-trivial dynamics. So, it makes sense from our point of view to generalize the technique of the reconstruction 
of potentials for the cases of closed and open universes also for tachyon fields.
For the tachyon field \cite{Sen} with the Lagrangian density 
\begin{equation}
L = -V(T)\sqrt{1-\dot{T}^2}
\label{Lagr}
\end{equation}
 the energy density and the pressure  are respectively 
\begin{equation}
\rho =\frac{V(T)}{\sqrt{1-\dot T^2}}
\label{en-tach}
\end{equation}
and
\begin{equation}
p=-V(T)\sqrt{1-\dot T^2} \hspace{0.1cm},
\label{pr-tach}
\end{equation}
while the equation of motion for the tachyon  is
\begin{equation}
\frac{\ddot T}{1-\dot T^2}+\frac{3\dot a\dot T}{a}+\frac{V_{\hspace{0.1cm},T}}{V}=0 \hspace{0.1cm},
\end{equation}
where $V_{\hspace{0.1cm},T} \equiv dV/dT$. Now we shall obtain an expression for the potential for which, under certain initial conditions for the tachyon field, the scale factor of the universe is precisely $a(t)$.  It follows from Eqs. (\ref{en-tach}) and (\ref{pr-tach}) that 
\begin{equation}
\dot T^2=\Big(\frac{\rho +p}{\rho }\Big) \hspace{0.1cm},
\label{dot-T}
\end{equation}
from which, with $T' \equiv \frac{\partial T}{\partial a}$ , we have
\begin{equation}
T'=\pm \frac{1}{a}\sqrt{\frac{\rho +p}{\rho\left(\rho-\frac{k}{a^2}\right) }},
\label{15}
\end{equation}
which yields, after integration,
\begin{equation}
T=T(a)=\pm \int \frac{da}{a}\sqrt{\frac{\rho(a)+p(a)}{\rho(a)\left(\rho(a)-\frac{k}{a^2}\right)}} \hspace{0.1cm}.
\label{tach1}
\end{equation}
Using now the relation
\begin{equation}
V=\sqrt{-\rho p}
\label{pot-tach1}
\end{equation}
and inverting (\ref{tach1}) we obtain the form of the tachyon potential:
\begin{equation}
V=V(a)=V[a(T)] \hspace{0.1cm}.
\end{equation}

In the fourth section we can consider also the case of the pseudotachyon fields \cite{Gorini:2003wa},
whose Lagrangian is 
\begin{equation}
L = V(T)\sqrt{\dot{T}^2-1}
\label{pseudo}
\end{equation}
and whose pressure is positive.
For pseudotachyon the formula for the potential is 
\begin{equation}
V(T) = \sqrt{\rho p}
\label{pseudo1}
\end{equation}
while the relation between the field and the cosmological factor $a$ is given again by the formula (\ref{tach1}).

\section{Reconstruction of scalar field potentials for closed and open universes}
In this section we consider both a closed and an open homogeneous and isotropic cosmological model filled with a barotropic fluid with the equation of state
\begin{equation}
p = w\rho,
\label{barotrop}
\end{equation}
where $w$ is a constant. We shall consider also the model of a universe filled with a mixture of cosmological constant and radiation.

As it easily follows from the energy conservation equation (\ref{conserv}) the dependence of the energy density on the scale factor is in this case 
\begin{equation}
\rho(a) = \frac{C_0}{a^{3(1+w)}},
\label{en-dep}
\end{equation}
where $C_0$ is a positive constant. Substituting the expressions (\ref{barotrop}) and (\ref{en-dep}) into Eq. (\ref{phi-int1}) 
we obtain
\begin{equation}
\varphi(a) =\pm \sqrt{1+w}\int \frac{da}{a\sqrt{1-\frac{k}{C_0}a^{1+3w}}}.
\label{integr1}
\end{equation}
Integrating Eq. (\ref{integr1}) one obtains 
\begin{equation}
\varphi(a) = \pm\frac{\sqrt{1+w}}{1+3w}\ln \frac{1+\sqrt{1-\frac{a^{1+3w}}{C_0}}}{1-\sqrt{1-\frac{a^{1+3w}}{C_0}}}
\label{integr-cl}
\end{equation}
for a closed universe,
\begin{equation}
\varphi(a) = \pm\frac{\sqrt{1+w}}{1+3w}\ln \frac{1+\sqrt{1+\frac{a^{1+3w}}{C_0}}}{\sqrt{1+\frac{a^{1+3w}}{C_0}}-1}
\label{integr-op}
\end{equation}
for an open universe and 
\begin{equation}
\varphi(a) = \pm \sqrt{1+w} \ln a
\label{integr-fl}
\end{equation}
for a flat universe. Inverting expressions (\ref{integr-cl})--(\ref{integr-fl}) and substituting the functions $a(\varphi)$ into Eq. (\ref{poten}) we obtain the explicit formulae for the potentials: 
 \begin{equation}
V(\varphi)=\frac{(1-w)}{2C_0^{\frac{2}{1+3w}}}\left(\cosh\frac{(1+3w)\varphi}{2\sqrt{1+w}}\right)^{\frac{6(1+w)}{1+3w}} 	
\label{pot-bar}
\end{equation}
for a closed universe,
\begin{equation}
V(\varphi )=\frac{(1-w)}{2C_0^{\frac{2}{1+3w}}}\left(\sinh \frac{(1+3w)\varphi }{2\sqrt{1+w}}\right)^{\frac{6(1+w)}{1+3w}} 
\label{pot-bar1}
\end{equation}
for an open universe and 
\begin{equation}
V(\varphi) = \frac{(1-w)C_0}{2}\exp\left(\pm \frac{3(1+w)\varphi}{\sqrt{1+w}}\right)
\label{flat}
\end{equation}
for a flat universe. 
Note that while for the closed and open universe the different choices of sign in the formulae (\ref{integr-cl}) and (\ref{integr-op}) give the same form of potential, in the flat case we have two different forms of the potential corresponding to the 
different choices of the sign in the right-hand side of Eq. (\ref{integr-fl}).

A universe filled with a mixture of the cosmological constant $\Lambda$ and radiation ($w = \frac13$) has the following energy density and pressure:
\begin{equation}
\rho(a) = \Lambda+ \frac{C_0}{a^4},\ \ p = -\Lambda +\frac{C_0}{3a^4}.
\label{mixt}
\end{equation}
 Correspondingly the formula (\ref{phi-int1}) looks now as 
\begin{equation}
\varphi(a) = \pm \sqrt{\frac{4C_0}{3}}\int \frac{da}{a\sqrt{\Lambda a^4-ka^2+C_0}}.
\label{mixt1}
\end{equation}
For the case of a closed universe the integration gives
\begin{equation}
\varphi(a) =\pm \frac{\sqrt{3}}{3}\ln\frac{2C_0-a^2+2\sqrt{C_0^2-C_0a^2+C_0\Lambda a^4}}{a^2},
\label{mixt2}
\end{equation}
for an open universe it is 
\begin{equation}
\varphi(a) =\pm \frac{\sqrt{3}}{3}\ln\frac{2C_0+a^2+2\sqrt{C_0^2+C_0a^2+C_0\Lambda a^4}}{a^2},
\label{mixt3}
\end{equation}
and for a flat universe it is 
\begin{equation}
\varphi(a) =\pm \frac{\sqrt{3}}{3}\ln\frac{C_0+\sqrt{C_0^2+C_0\Lambda a^4}}{a^2}.
\label{mixt4} 
\end{equation}
Inverting the expressions (\ref{mixt2})--(\ref{mixt4}) and substituting the explicit forms of $a(\varphi)$ into the general formula for the potential (\ref{poten}), one obtains the following formulae for the potentials:
\begin{equation}
V(\varphi) = \Lambda +\frac{C_0}{3}\left(\frac{\cosh^2 \frac{\sqrt{3}}{2}\varphi}{C_0^2}-\Lambda e^{\pm \sqrt{3}\varphi}\right)
\label{mixt5}
\end{equation}
for a closed universe,
\begin{equation}
V(\varphi) = \Lambda +\frac{C_0}{3}\left(\frac{\sinh^2 \frac{\sqrt{3}}{2}\varphi}{C_0^2}-\Lambda e^{\pm \sqrt{3}\varphi}\right)
\label{mixt6}
\end{equation}
for an open universe and 
\begin{equation}
V(\varphi) = \Lambda + \frac{\Lambda}{3}\sinh^2\left(\sqrt{3}\varphi \pm \frac12 \ln C_0\Lambda\right)
\label{mixt7}
\end{equation}
for a flat universe.

We see that the different choices of the sign in the expressions for the scalar field (\ref{mixt2}) and (\ref{mixt3})
give different forms of the potentials for the closed and open cosmological models 
in the formulae (\ref{mixt5}) and (\ref{mixt6}). 
In the case of the potential for the flat model (Eq. (\ref{mixt7})) the potentials corresponding to the different choice 
of sign for $\varphi(a)$  can be transformed one into another by a simple constant shift of the scalar field
(which is always defined up to an arbitrary constant of integration). Thus, in this case we have really only one form of the potential. 
It is easy to check that in the case when the cosmological constant is absent, the formulae (\ref{mixt5})--(\ref{mixt7}) 
are transformed into the formulae
(\ref{pot-bar})--(\ref{flat}), where the equation of state parameter is taken equal to $w = \frac13$.

\section{Reconstruction of tachyon and pseudo-tachyon potentials}
We begin this section with mentioning that the 
cases of a universe filled with a Chaplygin or anti-Chaplygin gas are trivial, because the tachyon potential has a constant value in these cases \cite{FKS}.
In fact, the equation of state for a Chaplygin gas is $p=-\frac{A}{\rho}$, with $A>0$ constant, and the expression for the potential is (\ref{pot-tach1}), from which we have
\begin{equation}
V(T)=\sqrt A
\end{equation}
which is constant.\\ For an anti-Chaplygin gas the situation is analogous, because the equation of state is $p=\frac{A}{\rho }$ and the expression for the potential is $V=\sqrt{p\rho }$, from which
\begin{equation}
V(T)=\sqrt A \hspace{0.1cm}.
\end{equation}
These are the simplest cases for the reconstruction of tachyon potentials. We then discuss other cases for which it is possible to found exact solutions.

Let us consider again the evolution of the universe driven by a perfect fluid with the equation of state (\ref{barotrop}). 
In this case the equation (\ref{tach1}) takes the form
\begin{equation}
T(a) = \pm \sqrt{1+w} \int \frac{da a^{\frac{1+3w}{2}}}{\sqrt{C_0-ka^{1+3w}}}.
\label{int-tach}
\end{equation}

 We consider some simple cases when this equation is integrable in elementary functions.
First we consider the case $w=-\frac23$, of the so called brane gas.
Then, integration of Eq. (\ref{int-tach}) gives
\begin{equation}
T(a) = \pm \frac{2\sqrt{3}}{3C_0}\sqrt{C_0a-k}.
\label{int-tach1}  
\end{equation}
Inverting this equation and using the formula (\ref{pot-tach1}). 
we come to 
\begin{equation}
V(T) = \frac{4\sqrt{6}C_0^2}{3(3C_0^2T^2 + 4k)}.
\label{pot-tach2}
\end{equation}

The second case is the string gas equation of state ($w = -\frac13$).
Now, the integration of (\ref{int-tach}) gives 
\begin{equation}
T(a) = \pm \sqrt{\frac{2}{3(C_0-k)}} a.
\label{int-tach3}
\end{equation}
Inverting the function $T(a)$ from the above equation and using the formula (\ref{pot-tach1}) we obtain 
\begin{equation}
V(T) = \frac{2}{3\sqrt 3}\frac{C_0}{(C_0-k)T^2}.
\label{tach-pot3}
\end{equation}

In the case of the pseudotachyons, when the equation of state parameter should be positive, we consider 
the case of the radiation $w = \frac13$. 
Now 
\begin{equation}
T = \pm \sqrt{\frac43}\frac{1}{k}\sqrt{C_0-ka^2}.
\label{int-tach4}
\end{equation}
This formula is valid if $k \neq 0$. The corresponding potential obtained by using the formula (\ref{pseudo1}) is 
\begin{equation}
V(T) = \frac{16\sqrt{3}C_0}{3(4C_0-3T^2)^2}.
\label{pot-pseud}
\end{equation}
Remarkably, the last formula does not depend on the sign of the spatial curvature of the universe.

For the sake of completeness, we shall give also the corresponding formulae for the case of a flat universe.
(This case was considered in detail in \cite{Padmanabhan:2002cp,Feinstein:2002aj,Gorini:2003wa}). 
Now the integral (\ref{int-tach}) has a simple form for an arbitrary $w$:
\begin{equation}
T(a) = \frac{2}{3\sqrt{(1+w)C_0}}a^{\frac{3(1+w)}{2}},
\label{flat5}
\end{equation}
while the corresponding potential for the tachyon and pseudotachyon fields is 
\begin{equation}
V(T) = \frac{4\sqrt{|w|}}{9(1+w)T^2}.
\label{flat10}
\end{equation}
Note that if we put in the formula (\ref{flat10}) the values $w=-\frac23$ or $w = -\frac13$ then this formula gives the same results as the formulae (\ref{pot-tach2}) and (\ref{tach-pot3}), where $k$ is taken equal to zero. Thus, the formulae 
(\ref{pot-tach2}) and (\ref{tach-pot3}) describe all the three cases: closed, open and flat. It is not so for the pseudotachyon potential expression (\ref{pot-pseud}) which is valid only for  closed and open universes.

\section{Conclusion}
In this Letter we have reconstructed the potential for a scalar field or a tachyon field, reproducing a given 
cosmological evolution in closed and open isotropic cosmological models, which, to our knowledge, were not considered before. We have obtained the explicit forms of such potentials for the cases of the evolutions driven by  a generic barotropic fluid and by radiation plus a cosmological constant, for the case of a scalar field, and by a Chaplygin gas, by an anti-Chaplygin gas and for other special cases,     for a tachyon field. Namely,  exact solutions for the tachyon potential have been obtained for some particular values of the barotropic index $w$, which for the tachyon field are $w=-\frac{2}{3}$ and $w=-\frac{1}{3}$, that are the cases of a brane gas and a gas of strings respectively, and for the pseudotachyon field $w=\frac{1}{3}$ (radiation). 

\section*{Acknowledgements}
This work was partially supported by the RFBR grant 11-02-00643.


\begin{thebibliography}{99}
\bibitem{Starobinsky:1998fr}
  Starobinsky, A.A.: JETP Lett. {\bf 68}, 757 (1998)
\bibitem{Burd:1988ss}
  Burd, A.B., Barrow, J.D.: Nucl. Phys. B {\bf 308}, 929 (1988)
\bibitem{Barrow} 
  Barrow,J.D.:
  Phys. Lett.  B  {\bf 235}, 40 (1990)
\bibitem{Ellis}
Ellis, G.F.R., Madsen, M.: Class. Quantum Grav. {\bf 8}, 667 (1991)
\bibitem{Zhuravlev:1998ff}
  Zhuravlev, V.M., Chervon , S.V., Shchigolev, V.K.:
  J. Exp. Theor. Phys. {\bf 87}, 223 (1998)
\bibitem{Yurov:2003zt}
  Yurov, A.:
  arXiv:astro-ph/0305019.
\bibitem{Yurov:2005bw}
  Yurov, A.V., Vereshchagin, S.D.:
  Theor. Math. Phys. {\bf 139}, 787 (2004)
\bibitem{Guo:2006ab}
  Guo, Z.K., Ohta, N., Zhang, Y.Z.:
  Mod. Phys. Lett. A  {\bf 22}, 883 (2007)
\bibitem{Guo:2005ata}
  Guo, Z.K., Ohta, N., Zhang, Y.Z.:
  Phys. Rev. D {\bf 72}, 023504 (2005)
\bibitem{inflation}
  Starobinsky, A.A.: Lect. Notes in Phys. {\bf 246}, 107 (1986);
  Linde, A.D.: Particle Physics and Inflationary Cosmology, Chur, Switzerland: Harwood (1990)
\bibitem{darkenergy}
  Sahni, V., Starobinsky, A.A.: Int. J. Mod. Phys. D  {\bf 9}, 373 (2000);
  Padmananbhan, T.: Phys. Rep. {\bf 380}, 235 (2003); Peebles, P.J.E., Ratra, B.: Rev. Mod. Phys. {\bf 75}, 559 (2003); Sahni, V.: Class. Quantum Grav. {\bf 19}, 3435 (2002);
 Copeland, E.J., Sami, M., Tsujikawa, S.: Int. J. Mod. Phys. D  {\bf 15}, 1753 (2006);
 Sahni, V., Starobinsky, A.A.: Int. J. Mod. Phys. D {\bf 15}, 2105 (2006)
\bibitem{acceleration}
 Riess, A. et al, Astron. J. {\bf 116}, 1009 (1998); Perlmutter,S.J. et al, Astroph. J. {\bf 517}, 565 (1998)  
\bibitem{Boisseau:2000pr}
  Boisseau, B., Esposito-Farese, G., Polarski, D., Starobinsky, A.A:
  Phys. Rev. Lett. {\bf 85}, 2236 (2000)
\bibitem{Andrianov:2007ua}
  Andrianov, A.A., Cannata, F., Kamenshchik, A.Y., Regoli, D.:
  JCAP  {\bf 0802}, 015 (2008)
\bibitem{Sen}
  Sen, A.:
  JHEP {\bf 0204}, 048 (2002)
\bibitem{Padmanabhan:2002cp}
  Padmanabhan, T.:
  Phys. Rev. D {\bf 66}, 021301 (2002)
\bibitem{Feinstein:2002aj}
  Feinstein, A.:
  Phys. Rev. D {\bf 66}, 063511 (2002)
\bibitem{Gorini:2003wa}
  Gorini, V., Kamenshchik, A.Y., Moschella, U., Pasquier, V.:
  Phys. Rev. D {\bf 69}, 123512 (2004)
\bibitem{Kam-Tron-Ven}
  Kamenshchik, A.Y., Tronconi, A., Venturi, G.:
  Phys. Lett. B {\bf 702}, 191 (2011)
\bibitem{we-Chap}
  Kamenshchik, A.Y, Moschella, U., Pasquier, V.: Phys. Lett. B {\bf 511}, 265 (2001)
\bibitem{Fabris}
  Fabris, J.C., Goncalves, S.V.B., de Souza, P.E.: Gen. Relativ. Gravit. {\bf 34}, 53 (2002); Bilic, N., Tupper, G.B., Viollier, R.D.: Phys. Lett. B {\bf 535}, 17 (2002); Bento, M.C., Bertolami, O., Sen, A.A.: Phys. Rev. D {\bf 66}, 043507 (2002) 043507; Gorini, V., Kamenshchik, A.Y., Moschella, U.: Phys. Rev. D {\bf 67}, 063509 (2003)
\bibitem{Carter}
  Carter, B.: Phys. Lett. B {\bf 224}, 61 (1989); Vilenkin, A.: Phys. Rev. D {\bf 41}, 3038 (1990)
\bibitem{k-ess}
Armendariz-Picon, C., Damour, T., Mukhanov, V.: Phys. Lett. B {\bf 458}, 209 (1999);
Armendariz-Picon, C., Mukhanov, V.F., Steinhardt, P.J.: 
Phys. Rev. Lett. {\bf 85}, 4438 (2000);
Phys. Rev. D {\bf 63}, 103510 (2001)
\bibitem{Born}
Born, M., Infeld, L.: Proc. R. Soc. Lond A {\bf 144},  425 (1934)
\bibitem{hydroK}
Diez-Tejedor, A.,Feinstein, A.: Int. J. Mod. Phys. D {\bf 14}, 1561 (2005)
\bibitem{FKS}
 Frolov, A.V., Kofman, L., Starobinsky, A.A.:
  Phys. Lett. B  {\bf 545}, 8 (2002)
\end{thebibliography}
\end{document}